\documentstyle[prb,twocolumn,aps,epsf]{revtex}

\tighten
\begin{document}
\draft
 
\title{Strong coupling phases of the $t-t'$ Hubbard model}

\author{J. Gonz\'alez}
\address{Instituto de Estructura de la Materia, CSIC, Serrano 123, 
28006 Madrid, Spain}
\author{J. V. Alvarez}
\address{Escuela Polit\'ecnica Superior, Universidad Carlos III,
Butarque 15, Legan\'es, 28913 Madrid, Spain}

\date{\today}
\maketitle
 
\begin{abstract}

We study the strong coupling regime of the $t-t'$ Hubbard model,
filled up to the level of the van Hove singularities, by means
of an exact diagonalization approach. We characterize the
different phases of the model by the different sectors of the
Hilbert space with given quantum numbers. By looking for the
ground state of the system, we find essentially the competition
between a state with incipient ferromagnetism and other
mimicking a $d$-wave condensate, which has the lowest
energy in a large region of the phase space.

\end{abstract}


During recent years there have been increasingly accurate
measures by angle resolved photoemission spectroscopy of
copper-oxide compounds, giving much insight into the
phenomenology of these materials\cite{photo}. 
Near the optimal doping for
superconductivity the hole-doped compounds use to show extended
van Hove singularities close to the Fermi level, which are
located near the high-symmetry points $(\pi , 0) , (0, \pi )$
\cite{photo,gofron}.
On the other hand, in the carrier free regime the materials show
antiferromagnetic correlations, with a dispersion relation which 
has peaks at the points $(\pm \pi /2, \pm \pi /2)$ \cite{wells}. 
A most
interesting problem is therefore to understand the drastic change 
that the Fermi surface may suffer by the influence of
doping\cite{wen}. 

The framework that has been proposed to address such theoretical
issues is that of the $t-t'-U$ model\cite{ttp} (or its strong coupling
version, the $t-t'-J$ model\cite{ttj}), 
as it is generally believed that
strong correlation effects have to be responsible for the
electronic properties of the cuprates. Next-to-nearest-neighbor
hopping $t'$ has to be introduced for a more accurate description of 
the dispersion relation in the insulating phase\cite{nazarenko}. 
The distinctive
feature of the $t-t'-U$ model is that the level of half-filling
does not coincide with the level corresponding to the two van
Hove singularities. Thus, it should be possible to establish a
clearer separation between the effects of the antiferromagnetic
correlations and those due to the appearance of the extended
saddle-points. There have been attempts to propose a purely
electronic mechanism of superconductivity in systems with van Hove
singularities close to the Fermi level\cite{newns,nos,levin}. 
What is essential in
those models is the existence of some enhanced channel favoring
the exchange of singlet pairs. They represent an alternative to
the picture earlier proposed in which the pairing interaction is
supposed to arise from the short-range antiferromagnetic
correlations\cite{afm}.

In the present paper we study the correlations which may
dominate at the van Hove singularities, by performing the exact
diagonalization of the $t-t'$ Hubbard model in a $4 \times 4 $
lattice (with periodic boundary conditions). The hamiltonian of
the model is
\begin{equation}
H = - t \sum_{i,j,\sigma} c^{+}_{i \sigma} c_{j \sigma} 
    + t' \sum_{i,j,\sigma} c^{+}_{i \sigma} c_{j \sigma} 
    +  U \sum_{i} n_{i \uparrow} n_{i \downarrow} 
\label{ham}
\end{equation}
where $\sigma = \uparrow , \downarrow$,
the first sum is over nearest neighbors $i,j$, 
the second sum over next-to-nearest neighbors,
and $n_i$ is the electron number operator at site $i$. 
A typical contour map of the dispersion relation (for $t'
< 0.5$ ) is shown in Fig. 1. We are specially interested in the
situation in which the van Hove shell, comprising the four
degenerate states at $(\pi , 0)$ and $(0, \pi )$, is
half-filled. We will pursue the determination of the lowest
energy state in each of the sectors with different quantum
numbers. By looking for the ground state of the model we will be 
able to study the interplay among the different
phases of the system and, as long as we are relying 
on a quantity (the ground state energy) 
that probes the lattice as a whole, we may hope to
predict properties with less influence from finite-size effects.
This procedure is similar to that applied to the Hubbard 
model\cite{tosatti}, where it appears to be safer than the
evaluation of correlation functions on the small lattice
scale\cite{fano1}.

\begin{figure}
\epsfysize=6cm\epsfbox{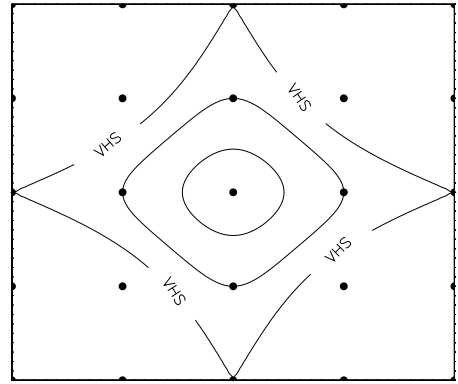}
\caption{Contour map of dispersion relation for the $t-t'$ Hubbard model.
         The van Hove shell (VHS) is shown.}
\end{figure}

Our starting point is the lattice in Fig. 1, filled with 12
particles for the 32 available one-particle states. For the
noninteracting theory ($U = 0$), this means that the ground
state is sixfold degenerate. The six states are given by the
different occupations of the van Hove points $A$ at $(\pi , 0)$
and $B$ at $(0, \pi )$. We may classify them according to the
total momentum ${\bf P}$ and total spin $S$, ending up with a 
$S = 1$ triplet at ${\bf P} = (\pi , \pi )$
\begin{eqnarray}
\left| T_1 \right. \rangle & = &  c^{+}_{A \uparrow} c^{+}_{B \uparrow}
\left| O \right. \rangle   \nonumber           \\
\left| T_0 \right. \rangle & = &  \frac{1}{\sqrt{2}} 
 \left( c^{+}_{A \uparrow} c^{+}_{B \downarrow} +
          c^{+}_{A \downarrow} c^{+}_{B \uparrow} \right)
\left| O \right. \rangle   \nonumber     \\
\left| T_{-1} \right. \rangle & = &  
   c^{+}_{A \downarrow} c^{+}_{B \downarrow}
\left| O \right. \rangle  
\label{ferr}
\end{eqnarray}
a spin singlet at ${\bf P} = (\pi , \pi )$
\begin{equation}
\left| P \right. \rangle  =   \frac{1}{\sqrt{2}} 
 \left( c^{+}_{A \uparrow} c^{+}_{B \downarrow} -
          c^{+}_{A \downarrow} c^{+}_{B \uparrow} \right)
\left| O \right. \rangle 
\end{equation}
and two spin singlets at ${\bf P} = (0,0)$
\begin{eqnarray}
\left| D \right. \rangle & = &  \frac{1}{\sqrt{2}} 
 \left( c^{+}_{A \uparrow} c^{+}_{A \downarrow} -
          c^{+}_{B \uparrow} c^{+}_{B \downarrow} \right)
\left| O \right. \rangle   \nonumber      \\
\left| S \right. \rangle & = &  \frac{1}{\sqrt{2}} 
 \left( c^{+}_{A \uparrow} c^{+}_{A \downarrow} +
          c^{+}_{B \uparrow} c^{+}_{B \downarrow} \right)
\left| O \right. \rangle 
\label{supr}
\end{eqnarray}
The states $\left| D \right. \rangle$ and $\left| S \right.
\rangle$ (as well as $\left| T_0 \right. \rangle$ and
$\left| P \right. \rangle$) differ in the quantum number
associated to the transformation by exchange of the two
components of the momentum, that is one of the generators of the
point symmetry group. The triplet state would be the precursor
of a state with nonzero magnetization for systems with larger
size, while $\left| P \right. \rangle $ would correspond to a
paramagnetic or antiferromagnetic state with no tendency to the
uniform alignment of spins. On the other hand, the
$\left| D \right. \rangle $ state mimics the fluctuation of
pairs between the two van Hove singularities, providing a
simplified version of a $d$-wave condensate. In what follows we
study how the degeneracy between these states is broken when the
interaction is turned on, interpreting the corresponding ground
state as an incipient signal of what should be the dominant
correlation in the model.

By considering the interaction of the particles in the van Hove
shell with the rest of closed shells in the Fermi sea (but without
allowing yet for particle-hole screening processes) we see that
the above ground state degeneracy is partially lifted. To first
order in $U$, the energy of each of the above multiplets is
\begin{eqnarray}
E_{\left| T \right. \rangle} & = & E_{kin} + \frac{U}{N}
  \left( \frac{n_{FS}}{2} + 2 \right) 
      \frac{n_{FS}}{2} + O(U^2)       \nonumber    \\
E_{\left| D \right. \rangle} & = & E_{kin} + \frac{U}{N}
  \left(  \left( \frac{n_{FS}}{2} + 1 \right)^2 
       - 1 \right)  + O(U^2)        \nonumber    \\
E_{\left| P \right. \rangle} & = & E_{kin} + \frac{U}{N}
  \left(  \left( \frac{n_{FS}}{2} + 1 \right)^2 
       + 1 \right)  + O(U^2)        \nonumber    \\
E_{\left| S \right. \rangle} & = & E_{kin} + \frac{U}{N}
  \left(  \left( \frac{n_{FS}}{2} + 1 \right)^2 
       + 1 \right)  + O(U^2)    
\end{eqnarray}
where $E_{kin}$ is the kinetic energy of the particles, $n_{FS}$
is the number of particles in closed shells and $N$ is the number
of lattice sites. We see that to this perturbative order the
states $\left| T \right. \rangle$ and $\left| D \right. \rangle$
have both the lowest energy. If we had an extended saddle point,
for instance, and the possibility of placing a number $m$ of
particles near $A$ or $B$ in a more degenerate level, it is
clear that aligning the spin of the $m$ particles would produce
a state with an energy $\approx E_{kin} + (U/N) \left(
n_{FS}/2 + m \right) n_{FS}/2 + O(U^2)$, while any state made of
a condensate of singlet pairs could hardly lower its energy from
$\approx E_{kin} + (U/N) \left( \left(
n_{FS} + m \right)/2 \right)^2 + O(U^2)$. It seems therefore
that, in the weak coupling regime of a model with a high
density of states close to the Fermi level, a state with a 
macroscopic amount of magnetization is energetically
favored\cite{mielke}. 
One has to bear in mind, however, that this argument works only
for weak coupling constant. The influence of particle-hole
processes and screening turns out to be crucial in the presence
of the van Hove singularities. In our $4 \times 4$ lattice
model one may check that, already at values of the coupling constant
as small as $U \approx 2t$, there is a value of $t'$ below which
the state corresponding to $\left| P \right. \rangle $ dressed
with particle-hole processes gets lower energy than that of 
$\left| T \right. \rangle $. Our exact diagonalization study
becomes relevant mainly in the strong coupling regime, where
there is no forecast of what may be the symmetry of the ground
state.  

As an illustration of the effects at large $U$ we have plotted
in Fig. 2 the evolution of the minimum values of the energy in
the different sectors with the respective quantum numbers of the
$\left| T \right. \rangle$, $\left| D \right. \rangle$ and
$\left| P \right. \rangle$ states, in the region of small $t'$
and $U = 8 t$. From the computational point of view, we have
reduced the Hilbert space in each case by using all the
generators of the point symmetry group\cite{fano2}, 
and we have implemented
a Davidson algorithm to obtain the lowest energy state in each
sector. The plot in Fig. 2 shows that above a certain value of
$t'$ the ground state is found in the sector with the symmetry
of the $\left| D \right. \rangle$ state and, in fact, it
corresponds to the original $\left| D \right. \rangle$ state in 
(\ref{supr}) dressed with a large amount of particle-hole
processes from the Fermi sea of our model. However, in the
region of very small $t'$ ($\leq 0.02 t$), the lowest energy is
reached in the sector of the $\left| P \right. \rangle$ state.
This is consistent with what we expect from the Hubbard
model, that is what we recover in the limit $t' \rightarrow 0$
although with a significant deviation from half-filling. The
ground state that we find in that regime corresponds to what
should be a state with paramagnetic or antiferromagnetic
properties in larger lattices. It is worthwhile to remark that
such state becomes a highly excited state in our study
for sufficiently large values of the next-to-nearest neighbor
hopping, being also overriden by the lowest energy state with
spin $S = 1$ above some value of $t'$. The two crossings of
levels that we appreciate in Fig. 2 have a genuine quantum
character, since we have already seen that at very weak coupling
constant the ground state has to become essentially either the 
triplet state in (\ref{ferr}) or the $\left| D \right. \rangle$
state in (\ref{supr}).
\begin{figure}

\epsfysize=6cm\epsfbox{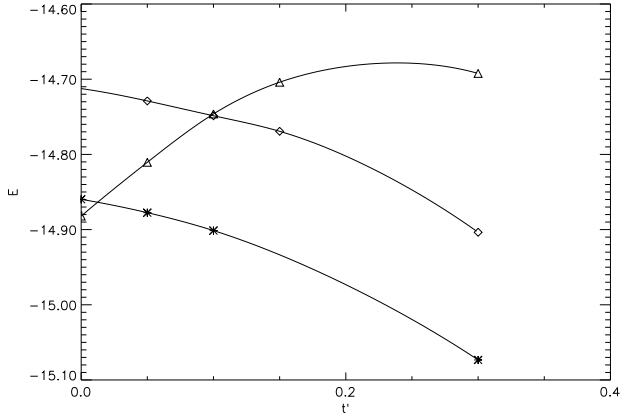}
\caption{Minimum value of energy in the sectors 
         of $\left| D \right. \rangle$,$\left| T \right. \rangle$ and 
        $\left| P \right. \rangle$ states (asterisks,diamonds and triangles
         resp.)
         in the region of small  $t'$ 
        and $U=8t$. }
\end{figure}

We have also studied the behavior of the model at increasing values
of $U$ and next-to-nearest neighbor hopping $t' = 0.3 t$, that is 
adequate for making contact with the phenomenology of the
cuprates. The sectors which may have candidates for the ground
state of the system have again the quantum numbers of 
$\left| T \right. \rangle$, $\left| D \right. \rangle$ and 
$\left| P \right. \rangle$, at least up to $U = 12 t$. 
It can be checked that states with momentum different from
$(0,0)$ or $(\pi , \pi )$, or spin different from 0 or 1,
have much higher energy. Furthermore, any state with the quantum
numbers of $\left| S \right. \rangle$ is always an excited state
of the system. In the mentioned range of the coupling constant,
the ground state turns out to be the 
$\left| D \right. \rangle$ state conveniently dressed by
particle-hole processes from the closed-shells of the Fermi sea. We
have represented in Fig. 3 the difference in energy of such
state with the state of minimum energy in the sector of 
$\left| P \right. \rangle$, and in Fig. 4 with respect to the
corresponding lowest energy state built from $\left| T \right.
\rangle$. It 
becomes clear that the state which could signal the appearance
of antiferromagnetic order has no chance of being the ground
state of the system, for the value of $t' = 0.3t$ considered.
Moreover, the results plotted in Fig. 4 show that the state
which mimics the $d$-wave condensate has always lower energy than
the triplet state up to the largest value $U = 10t$ considered.
It is important to note that at the largest values of $U$ the
energy difference between the two levels has a very slow decline,
what makes unlikely that the triplet state may become the ground 
state for a near value of $U$. The situation seems similar to the
case of Nagaoka ferromagnetism, where this effect is found in the
small $4 \times 4$ lattice for very large values of $U$ 
( $> 60t$ ), although the nature of the ferromagnetism may be 
quite different in the present model. The persistence of the 
$\left| D \right. \rangle $ state as the ground state of the system over 
a large region of the parameter space provides good reason to
believe that the corresponding $d$-wave condensate may continue
playing that role in the larger lattices. 

\begin{figure}
\epsfysize=6cm\epsfbox{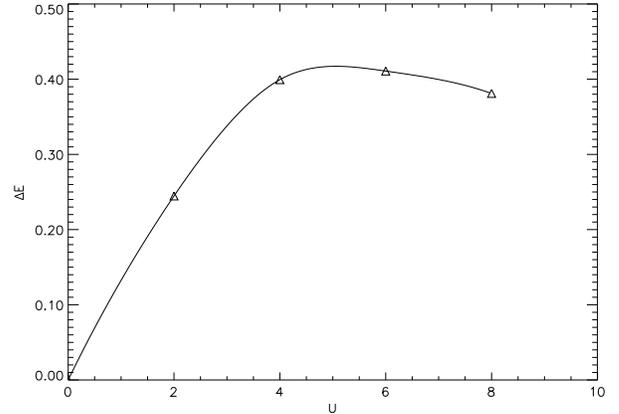}
\caption{ Difference between minimum energies of the sectors containing 
        $\left| D \right. \rangle $  and $ \left| P \right . \rangle $ 
        states (t'=0.3).}
\end{figure}

\begin{figure}
\epsfysize=6cm\epsfbox{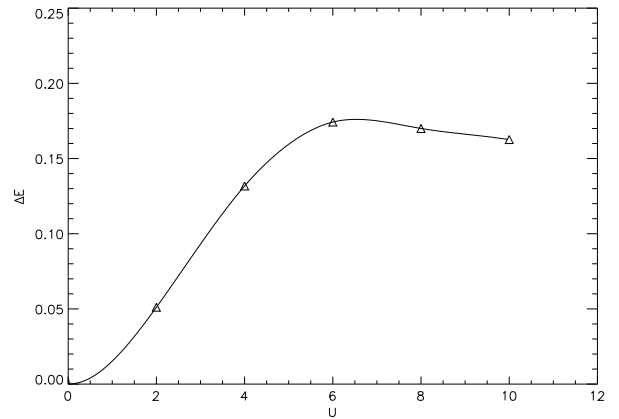}
\caption{Difference between minimum energies of the sectors containing 
        $\left| D \right. \rangle $  and $ \left| T \right . \rangle $ 
        states (t'=0.3).}
\end{figure}

The possible crossing of levels between the singlet
with $d$-wave symmetry and the $S = 1$ state is also reminiscent of a 
phenomenon studied in small clusters of the hexagonal lattice. In 
those systems the regime in which the singlet prevails as the 
ground state has been linked to the effect of pair binding above 
half-filling. The hexagonal lattice at half-filling has strictly 
two Fermi points, what makes likely that such pair binding may
rely on the resonance of electron pairs at the Fermi level. The 
important difference with respect to the model with the two van
Hove singularities is that the density of states for the hexagonal
lattice goes to zero at the Fermi points, which is just the opposite 
of the situation that we are facing. This may explain why the ground
state with pairs fluctuating between the two van Hove points is more
stable in this case, persisting over a much wider range of values of
$U$ than in the clusters studied in Ref. \onlinecite{white}.

The results that we have obtained have to be 
taken anyhow with the
reserve inherent to the use of a small lattice in the exact
diagonalization approach. As we have pointed out before, our
analysis is likely to be relevant in the study of the strong
coupling regime, where the effects of screening due to
particle-hole processes are difficult to assess by any other
method. Moreover, our results apply directly to the zero
temperature regime of the system, while the use of other
approaches like the quantum Monte Carlo method make very
difficult to perform extrapolations to the limit of vanishing
temperature. This drawback seems to be overcome anyhow in the
study of the $t-t'$ Hubbard model of Ref. \onlinecite{sorella}, 
where it is
claimed that at $U = 4 t$ and $t' = 0.47 t$ a signal of
ferromagnetism is present in the model. For those values of the
couplings our picture does not differ much qualitatively
from what is shown in
Figs. 3 and 4, but this is consistent with the fact that the
signal found in Ref. \onlinecite{sorella}
is so weak that requires lattices as
large as $16 \times 16$ for the ferromagnetism to be observed. 
It has to be stressed that the value $t' = 0.47t$ is
very close to the point in which the Fermi sea degenerates into
a pair of straight lines, making the analysis of the scattering
processes of Ref. \onlinecite{sorella} 
in terms of ladder diagrams quite
appropriate. Our approach, however, has to be relevant for a
completely different regime, that arises when particle-hole
processes induce strong screening effects in the model\cite{mc}.

Of course, in an ideal analytical framework one should be able to deal
simultaneously with the description of screening processes and
scattering in the particle-particle channel. One way to
accomplish this would be to adopt a renormalization group
approach, taking first into account the scaling of the different
channels and studying then the behavior of the response
functions. The two channels that are not irrelevant, apart from
the forward scattering channel\cite{shankar}, 
are the scattering $V$ of 
pairs about the same van Hove singularity and the scattering
$\tilde{V}$ of pairs from one singularity to the other. The
regime in which the intra-singularity screening dominates making
$V < \tilde{V}$ (what happens above certain value of $t'$) seems
to lead to an instability favoring the condensation of pairs
with opposite amplitude in the two van Hove singularities\cite{nos}. 
This is
nothing but a new version of the Kohn-Luttinger mechanism\cite{kl}, in
which the anisotropy of the screening effects arise in a natural
way. The superconducting instability may therefore be competing
with the above mentioned ferromagnetic instability, at least for
values of $t'$ not very close to $0.5t$. It would be worthwhile
to develop some computational scheme with which to perform an
extrapolation of our results to larger lattice sizes, in order
to see whether the evidence found for $d$-wave superconductivity
survives in the strong coupling regime.

We want to thank F. Guinea, R. Hlubina and S. Sorella for
useful discussions during the development of this work.

\end{document}